\documentclass[pre,aps,twocolumn]{revtex4}
\usepackage{graphicx,epsf,subfigure,amsbsy,amsmath,amsfonts,float}
\usepackage[latin1]{inputenc}

\begin{document}
\title{Embeddings of low-dimensional strange attractors:\\
Topological invariants and degrees of freedom}

\author{Nicola Romanazzi$^1$, Marc Lefranc$^2$, and Robert Gilmore$^1$} 

\affiliation{$^1$Physics Department, Drexel University,
Philadelphia,  Pennsylvania 19104, USA}

\affiliation{$^2$Laboratoire de Physique
  des Lasers, Atomes, Mol\'ecules, UMR CNRS 8523, Centre d'\'Etudes et
  de Recherches Lasers et Applications, Universit\'e des Sciences et
  Technologies de Lille, F-59655 Villeneuve d'Ascq, France}

\date{\today, {\it Physical Review E}: To be submitted.}

\begin{abstract}
When a low dimensional chaotic attractor is embedded in
a three dimensional space its topological properties are
embedding-dependent.  We show that there are just three
topological properties that depend on the embedding:
parity, global torsion, and knot type.
We discuss how they can change with the embedding.
Finally, we show that the {\it mechanism} that is 
responsible for creating chaotic behavior is an
invariant of all embeddings.  These results apply only
to chaotic attractors of genus one, which covers the
majority of cases in which experimental data have been
subjected to topological analysis.  This means that
the conclusions drawn from previous analyses, for example
that the mechanism generating chaotic behavior is a
Smale horseshoe mechanism, a reverse horseshoe, a
gateau roul\'e, an $S$-template branched manifold, $\dots$,
are not artifacts of the embedding chosen for the analysis.
\end{abstract}


\maketitle

\section{Introduction}
\label{sec:introduction}

Chaotic time series have been generated by a large number
of experiments.  Typically a scalar time series is available,
and a chaotic attractor must be generated from the scalar time
series using some embedding procedure.  The algorithm of choice is
the time delay embedding \cite{Whitney, Packard, Takens}
although differential embeddings, Hilbert transform
embeddings, and singular value decomposition embeddings
have also been used \cite{RMP, ToC}.

The properties of embedded chaotic attractors have been analyzed 
along three distinct mathematical lines: geometric, dynamical, 
and topological.  Geometric analyses involve computing various fractal
dimensions \cite{fractals}.  Dynamical analyses involve computing Lyapunov
exponents and the average Lyapunov dimension \cite{lyaps}.  
Topological analyses concentrate on the global topological properties
of an attractor by studying how stretching and squeezing
mechanisms organize the unstable periodic orbits embedded in
the attractor \cite{RMP, ToC, RRR, classification, Gabo}.  

In all approaches, it is assumed that the 
embedding adopted creates a diffeomorphism
between the underlying (invisible) experimental attractor that generates the
data and the embedded, or reconstructed, chaotic attractor
\cite{Packard,Takens}.  
Since the geometric and dynamical measures (dimensions and exponents)
are invariant under diffeomorphisms, in principle these real numbers
are embedding-independent.  In practice they are difficult to
compute, and become increasingly difficult to compute 
as the length of the time series and/or the 
signal to noise ratio decreases.
Further, there is no independent way to compute errors 
for the estimates of these real numbers.
It was even shown in \cite{Lef92} that in some experimental data sets
estimates of the fractal dimensions are diffeomorphism-dependent.  
Spurious Lyapunov exponents occur when the embedding dimension
exceeds the dimension of the dynamical system.
This has been addressed in \cite{Brown91,Sauer98} but remains
an open problem.

By constrast, topological analyses 
on three dimensional embeddings have been carried out with
relatively short experimental data sets and are robust against
noise.  In addition, this analysis method is overdetermined.
The stretching and squeezing mechanism creating the embedded
chaotic attractor can be determined from a small number
of unstable periodic orbits and used to predict the
topological organization of all remaining orbits.  These
predictions (linking numbers, relative rotation rates) 
either agree or do not agree
with those for orbits extracted from the embedded
chaotic attractor.  In the latter case the model describing
stretching and squeezing must be rejected.

What has never been satisfactorily understood is the
relation between the topological properties of the
underlying (invisible) experimental attractor that generates the
data and the chaotic attractor that has been constructed
through an embedding of the data.  We illustrate this
difficulty with two examples.  (1) The Lorenz attractor 
\cite{Lor63} is
described by variables $(x(t),y(t),z(t))$.  One three-dimensional
embedding is the obvious one: $(X_1,X_2,X_3)=(x,y,z)$.
The chaotic attractor in this embedding is invariant
under rotations: $(X_1,X_2,X_3) \rightarrow (-X_1,-X_2,X_3)$.  
On the other hand, if a single variable is observed 
(either $x(t)$ or $y(t)$) \cite{Bau91}
the chaotic attractor created through the differential
embedding $(X_1,X_2,X_3)=(x,\dot{x},\ddot{x})$ will
exhibit inversion symmetry
$(X_1,X_2,X_3) \rightarrow (-X_1,-X_2,-X_3)$ \cite{Let96, SoC06}.
(2) The chaotic behavior of B\'enard-Marangoni
fluid convection in a square cell \cite{Ond93} was modeled
by a periodically driven Takens-Bogdanov nonlinear oscillator
\cite{Hue96}.  This model
was studied using a time delay mapping of the form
$(X_1,X_2,X_3)=(x(t),\dot{x}(t),x(t-\tau))$ \cite{Min95}.
For a range of values of the time delay $\tau$,
$\tau_1 < \tau < \tau_2$, the image of the data under this
mapping exhibits self intersections and the mapping
is therefore not an embedding (technically, it is an immersion)
\cite{Min95, Aruna}.  
For $\tau < \tau_1$ and $\tau_2 < \tau$ the mapping is an embedding.
The topological organization of the unstable periodic
orbits under the two embeddings is different, so the
global topological structure of the two embedded attractors
is not equivalent \cite{Aruna}.

This discussion brings us to the crucial question:
When topological information about a chaotic attractor
is determined from a three-dimensional embedding of the chaotic attractor,
what part of that information is embedding-dependent and
what part is embedding-independent?  In this work we
answer this question for a large class of chaotic attractors.
These consist of all chaotic attractors of ``genus-one'' type
\cite{TTRG1, TTRG2}: their natural phase space is 
equivalent to a torus. This includes the chaotic attractor
discussed in the example (2) above. The answer is that
the ``mechanism'' (defined in Sec. \ref{sec:mechanisms} below)
is independent of embedding. Further, the topological organization
of all periodic orbits in the attractor 
can differ in a very limited number
of ways (parity, global torsion, and knot type, see
Sec. \ref{sec:embeddings} below).
This crucial question remains open for chaotic attractors whose
natural phase space is a torus of genus $g$ ($g > 1$).
This includes the Lorenz attractor as well as many other
chaotic attractors \cite{TTRG1, Aziz-Alaoui}.  It also
remains open for all higher-dimensional (hyper-)chaotic
attractors.

\section{Assumptions}
\label{sec:assumptions}

We make the following assumptions:

\begin{enumerate}
  \item  A deterministic process (e.g., laser equations,
Navier-Stokes equations) acts to generate an  experimental 
chaotic attractor that is three-dimensional.  
A single variable  (e.g., laser intensity, fluid surface height) 
is measured.
  \item  At least one embedding of this scalar time series 
in $\mathbb{R}^3$ can be constructed.  This embedding
creates a diffeomorphism between the  original experimental
chaotic attractor  and the embedded  or ``reconstructed'' 
chaotic attractor.
  \item  The embedded chaotic attractor is of genus-one type: 
that is, it can be enclosed in a genus-one bounding torus 
\cite{TTRG1, TTRG2}.
\end{enumerate}

Some remarks about these assumptions are in order. We assume in (1)
that there is an experimental chaotic attractor 
and that it is three dimensional.  By three-dimensional we mean
explicitly that there is a three dimensional manifold in the
phase space that contains the attractor.  We require this assumption
on dimension because, at the present time, topological analysis
methods based on templates are only applicable to three
dimensional chaotic attractors, that is, those that that exist in
three-dimensional manifolds.
The assumption that the deterministic process generates
a low-dimensional attractor is also strong: the Navier-Stokes
and the full laser equations are partial differential equations
rather than sets of ordinary differential equations,
and act in Hilbert spaces rather than finite dimensional
phase spaces \cite{Sal62}.

Assumption (2) is necessary because the Whitney embedding 
theorem \cite{Whitney} and its dynamical variants  
\cite{Packard, Takens} 
only guarantee that the three-dimensional manifold containing
the chaotic attractor can be embedded into a space of 
sufficiently high dimension ($6=2 \times 3$), 
but do not ensure that it can be done into a three-dimensional 
phase space.
In practice,
whether this assumption holds can be tested {\em a posteriori} by verifying
that the topological invariants measured are consistent with
a single two-dimensional branched manifold.
The diffeomorphism property that is assumed
of the mapping is the standard assumption for all approaches  
to analysis of embedded data \cite{Packard, Takens}.

Assumption (3) is crucial for our result.  It allows us to
reduce the problem of the inequivalence of embeddings of chaotic
attractors to the problem, already solved \cite{Rolfsen}, of the 
equivalence classes of diffeomorphisms of the solid torus into the
three-dimensional euclidean space $R^3$.
In the higher genus case (e.g., Lorenz attractor) the spectrum
of inequivalent diffeomorphisms (embeddings) of these attractors is
related to the spectrum of inequivalent diffeomorphisms
of the higher genus tori to themselves, which remains to
be studied.

\section{Preliminary Remarks}
\label{sec:preliminary}

We begin by recalling that diffeomorphisms map periodic
orbits to periodic orbits.  If ${\bf x}(t)$ is a point
on a periodic orbit so that ${\bf x}(t+T)={\bf x}(t)$, 
then under a diffeomorphism that takes ${\bf x \rightarrow y}$,
${\bf y}(t)={\bf y}(t+T)$.  This means that periodic 
orbits are neither created nor annihilated by diffeomorphisms.  
In particular, the spectrum of periodic orbits associated
with (``in'') a chaotic attractor is an invariant of
diffeomorphisms.  On the other hand their topological
organization, as encoded by their topological invariants
(linking numbers, relative rotation rates) could change
under diffeomorphism.

We will describe exactly how these topological invariants
can change under diffeomorphism when the phase space containing the chaotic
attractor is a torus $D^2 \times S^1$, where $D^2$
is a disk in the plane ($D^2 \subset R^2$) and $S^1$
is parameterized by $\phi$, usefully considered as a
phase angle mod $2\pi$.  In this phase space trajectories
can be expressed in the form $(x(t),y(t),\phi(t))$.
This class includes nonautonomous dynamical systems such
as the periodically driven Duffing, van der Pol, and
Takens-Bogdanov nonlinear oscillators where 
$\phi$ and $t$ are linearly related, and autonomous
dynamical systems whose phase space projection
$(x,\dot{x})$ exhibits a `hole in the middle'
(e.g., R\"ossler system \cite{Ros76} at $(a,b,c)=(0.398,2.0,4.0)$).
It includes other autonomous dynamical systems with
a hole in the middle that is present but obscured
by simple projections (e.g., R\"ossler system
at $(a,b,c)=(0.398,2.0,13.3)$).  For this class of systems
the phase $\phi = \phi(t)$ is a monotonic function of 
the time $t$.  This discussion explicitly excludes
attractors of genus $g \ge 2$ with two or more
`holes in the middle', such as the Lorenz attractor.

In the work to follow we seek a discrete enumeration of embeddings,
or diffeomorphsims, of strange attractors.  To achieve this end
it is necessary to ``mod out'' continuous degrees of freedom
associated with diffeomorphisms.  To do this we introduce the idea
of isotopy.  Two embeddings $f_0$ and $f_1$  are isotopic is
there is a one parameter family of mappings, $f(s)$, with
$f(0) = f_0$, $f(1) = f_1$ and $f(s)$ is an embedding for 
{\em all} $s$, $0 \le s \le 1$.  Such a family of embeddings
merely deforms the phase space smoothly.  The topological organization
of periodic orbits is unchanged under isotopy.  For if two orbits
intersected during the deformation from $s=0$ to $s=1$ the uniqueness
theorem would be violated and the mapping $f(s)$ (for some $s$)
would not be
a diffeomorphism.  For this reason isotopic mappings are in some
sense equivalent.  The sense is that all topological indices for
orbits in a strange attractor are the same for all embeddings
in the same isotopy class.

Our problem therefore reduces to 
(1) classifying the set of isotopy classes of diffeomorphisms 
$D^2 \times S^1 \rightarrow D^2 \times S^1$,
(2) classifying the set of isotopy classes of diffeomorphisms 
$D^2 \times S^1 \rightarrow R^3$, and
(3) determining how topological invariants change from one
class to another.  The first two parts of this program are
resolved in Secs. IVa and IVb.  The third part is discussed in
Secs. V and VI.  A more detailed exposition of these points is
presented in Appendix A.

\section{Embeddings of a Torus}
\label{sec:embeddings}

Diffeomorphisms of the torus fall into
two broad classes: intrinsic and extrinsic
\cite{Rolfsen, Weeks}.
Intrinsic diffeomorphisms are mappings
of the torus to itself ``as seen from the inside.''
Specifically, they are mappings $D^2 \times S^1
\rightarrow D^2 \times S^1$.  Extrinsic diffeomorphisms
describe how the torus sits in $R^3$.  They are
mappings $D^2 \times S^1 \rightarrow R^3$.
Intrinsic diffeomorphisms are responsible for two
of the three degrees of freedom mentioned in the
abstract and introduction: parity and global torsion.
Extrinsic diffeomorphisms are responsible for the 
first two and in addition the third:  knot type.

\subsection{Intrinsic Diffeomorphisms.}  These also fall into 
two classes: those that are isotopic to the identity
and those that are not.  

{\bf Isotopic to the Identity.}  Diffeomorphisms that 
are isotopic to the identity smoothly deform the phase 
space.  Therefore they do not change the topological organization
of the periodic orbits in it.  Under these
diffeomorphisms the topological invariants of 
periodic orbits remain unchanged.  

\begin{figure}[htb]   
\begin{center}
\includegraphics[angle=0,width=7.0cm]{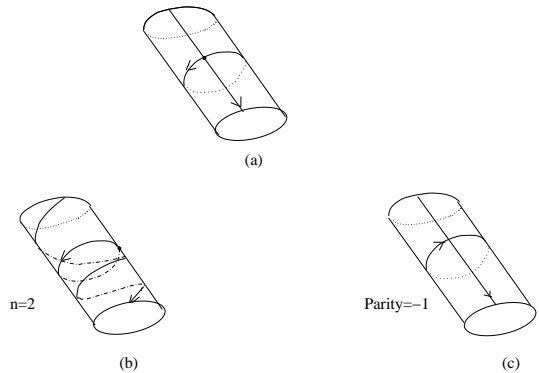}
\end{center}
  \caption{(a) Two nonisotopic circles are drawn on
the surface of the solid torus containing a
chaotic attractor.  The longitude is
oriented along the direction of the flow.
The meridian is oriented by the right hand rule.
The solid torus is mapped diffeomorphically to
a torus with (b) $n=2$  or (c) negative parity.}
  \label{fig:tori}
\end{figure}

{\bf Not Isotopic to the Identity.} Mappings of the torus
to itself that are not isotopic to the identity
have been classified \cite{Rolfsen}.  The idea is
as follows.  On the two-dimensional surface
$T^2 = \partial (D^2 \times S^1)$ that is the boundary
of the solid torus it is possible to construct two
circles that cannot be deformed to a point,
as shown in Fig. \ref{fig:tori}.  
We orient both. The longitude is oriented
along the direction of the dynamical system flow.
The meridian bounds a disk that can be used as a
Poincar\'e surface of section.  It is oriented according
to the right hand rule. Up to isotopy
(the class of diffeomorphisms considered in the preceeding
paragraph) the inequivalent diffeomorphisms of the torus to
itself are classified by their action on the longitude and
meridian by the matrix \cite{Rolfsen}

\begin{equation}
\left[  \begin{array}{cc}
1 & n \\ 0 & \pm 1 \end{array} \right]
\label{finite_diffeo}
\end{equation}
The integer $n$ describes the number of rotations of the
longitude about the core (center line) of the torus
as the phase angle $\phi$ increases from 0 to $2\pi$.
The integer $\pm 1$ indicates whether the diffeomorphism
preserves or reverses the orientation of the meridian.
We identify $\pm 1$ with parity and $n$ with global torsion
in Sec. \ref{sec:mechanisms}.

{\bf Remark.}  The matrices presented in Eq.(\ref{finite_diffeo})
are group operations.   Diffeomorphisms of the torus to itself
form a group.  The subset that is isotopic to the identity
forms a subgroup that is invariant in the larger group.
The quotient of these two groups therefore forms a group.
This group is discrete.  It is generated by two operations,
represented by the matrices

\begin{equation}
\left[  \begin{array}{cc}
1 & 1 \\ 0 &  1 \end{array} \right]
~~~~~~~{\rm and}~~~~~~~
\left[  \begin{array}{cc}
1 & 0 \\ 0 & -1 \end{array} \right]
\label{cosets}
\end{equation}
The first describes the generator that produces a
uniform rotation along the axis of the torus:
$((x+iy),\phi) \rightarrow ((x+iy)e^{i \phi}, \phi)$.  
The second
generator produces the effect of looking into a
mirror: $(x,y,\phi) \rightarrow (x,- y ,\phi)$.
This coset decomposition says simply that every intrinsic
diffeomorphism can be constructed by composing a
diffeomorphism isotopic to the identity with
one from the discrete group whose matrix representation
is given in Eq.(\ref{finite_diffeo}).

\subsection{Extrinsic Diffeomorphisms.}

\begin{figure}[htb]   
\begin{tabular}{c}
\includegraphics[angle=0,width=7.0cm]{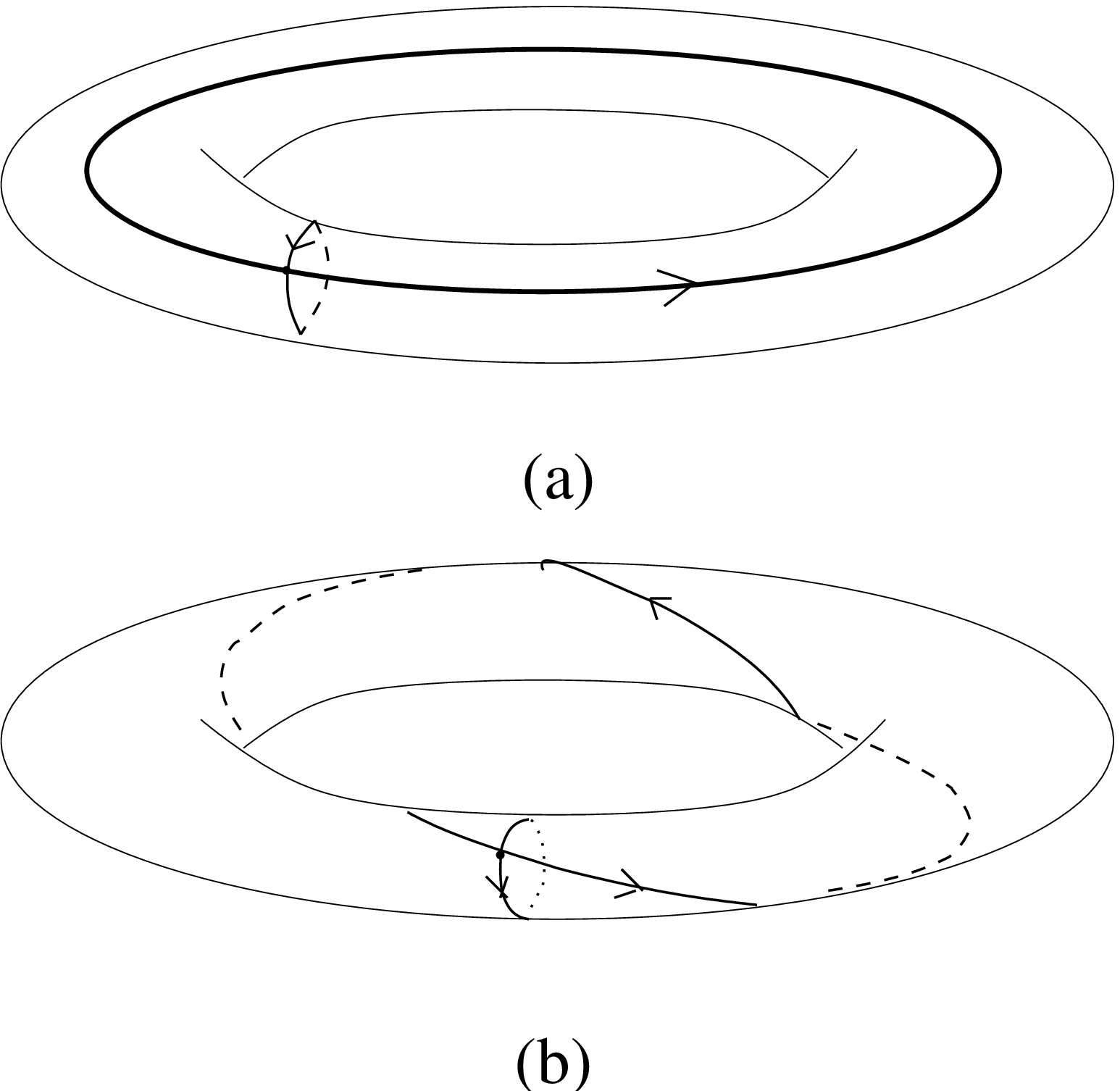}\\
 \\
\includegraphics[angle=0,width=6.0cm]{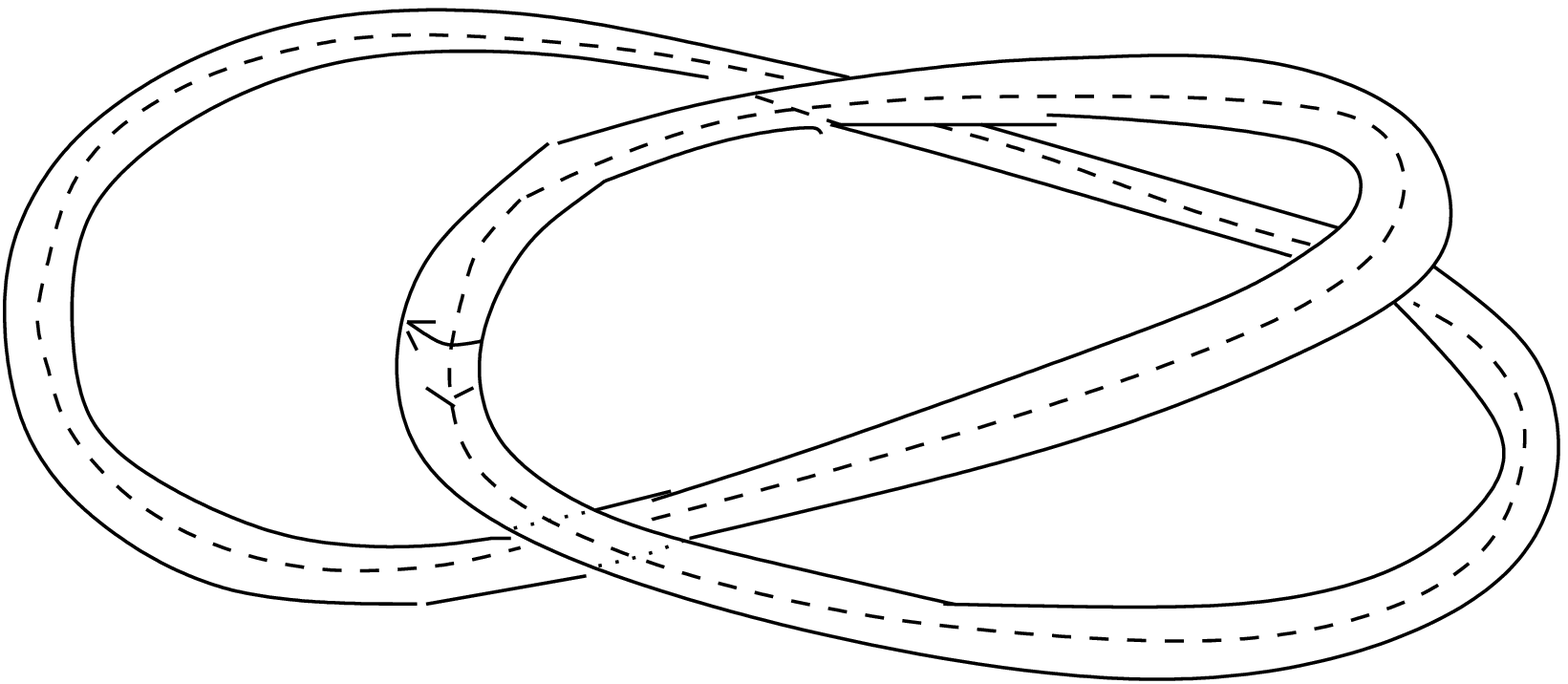}
\end{tabular}
  \caption{(a) The torus $D^2 \times S^1$ of Fig. \ref{fig:tori}(a)
is embedded in a natural way in $R^3$.  In this embedding the
core of the torus is a circle of radius $R$.
the torus can also be mapped
into $R^3$ with a nonzero framing index $f$.
The framing index is $+1$ in the embeddings (b)
and (c).}
  \label{fig:embed_frame}
\end{figure}

The mapping of $D^2 \times S^1$ into $R^3$ shown in Fig.
\ref{fig:embed_frame}(a) is called the `natural embedding'
\cite{Rolfsen}. 
One natural embedding of a chaotic
attractor with coordinates $(x_1(\phi),x_2(\phi),
\phi)$ in $D^2 \times S^1$ into $R^3$ is $(X(t),Y(t),Z(t))$,
with $t = \phi$ and $X=(R - x_1)\cos \phi$,
$Y=(R-x_1)\sin \phi$, and $Z=x_2$.  This is an
embedding provided the circle is ``bigger'' than
the attractor.  Specifically, if the radius of the
disk $D^2$ containing the attractor is $a$, so that
$x_1^2(\phi)+x_2^2(\phi) < a^2$ for all $\phi$,
then $R>a$ guarantees that no self-intersections
occur in the natural embedding.

The circle is the simplest knot in $R^3$.
Other knots in $R^3$ can be used as central curves
for other extrinsic embeddings.  The knot
${\cal K}$ has coordinates 
${\bf K}(\phi)=(K_1(\phi),K_2(\phi),K_3(\phi))$ with
${\bf K}(\phi)={\bf K}(\phi+2\pi)$.  As with any smooth
space curve \cite{Struik} this knot has a moving
coordinate system ({\it repere mobile}) with orthogonal unit vectors
${\bf t}(\phi),{\bf n}(\phi), {\bf b}(\phi)$.  The section of a
chaotic attractor in $D^2 \times S^1$ at phase angle
$\phi$ is lifted into the plane in $R^3$ perpendicular to
the tangent vector ${\bf t}(\phi)$ at ${\bf K}(\phi)$ by
the mapping $(x_1(\phi),x_2(\phi),\phi) \rightarrow
{\bf K}(\phi) + x_1(\phi){\bf n}(\phi)+x_2(\phi){\bf b}(\phi)$.
This mapping is an embedding provided there are no self
intersections.  This is guaranteed provided two conditions
are satisfied \cite{Per05}:

\begin{description}
  \item[Local condition:]  The radius of curvature
of ${\cal K}$ is everywhere greater than $a$.
  \item[Global condition:]  The curve ${\cal K}$ is ``big enough.''
This means specifically that all nonzero local minima of
$|{\bf K}(\phi_1)-{\bf K}(\phi_2)|$ are larger than $2a$.
\end{description}

An important integer is associated with each knot
${\cal K}$.  This is its framing index, $f$
\cite{Rolfsen}.  It describes how many
times the vectors ${\bf n}$ and ${\bf b}$ wind around
${\bf t}$ as the knot is traversed.  Specifically, it is
the gauss linking number of two closed curves
in $R^3$.  One closed curve is the knot itself.
The other is obtained by displacing it a small distance
along the normal vector.  Its coordinates are given by setting 
$(x_1(\phi), x_2(\phi),\phi)=(1,0,\phi)$ in the mapping above.
We use this integer in Sec. \ref{sec:topological} to describe
the problems of the delay embeddings of the fluid data
presented in Sec. \ref{sec:introduction}, Example (2)
(embedding of Benard-Marangoni fluid data).  Embeddings of the torus
into $R^3$ with framing index $f=+1$ are shown in Fig.
\ref{fig:embed_frame}(b) and (c).

{\bf Remark.}  As Fig. \ref{fig:embed_frame} shows, choice
of a knot in $R^3$ for the center curve of the embedded torus
is independent of the choice of the framing index
of the embedded torus. The knot type of the center curve is one degree
of freedom of embeddings of a genus-one torus into $R^3$.
Two other degrees of freedom, the framing index (which is equivalent
to global torsion) and parity have already been
encountered in diffeomorphisms $D^2 \times S^1
\rightarrow D^2 \times S^1$.

{\bf Remark.}  It is pedantically more
accurate to describe extrinsic embeddings as diffeomorphisms
$D^2 \times S^1 \rightarrow D^2 \times S^1 \subset R^3$.
In the remainder we forgo this mathematical precision.

\section{Mechanisms}
\label{sec:mechanisms}

Chaotic attractors in three dimensional spaces are
characterized by the spectrum and topological organization
of their unstable periodic orbits (UPOs) \cite{RMP, ToC}.  
The topological 
organization of the periodic orbits is summarized by a
knot holder (also called a branched manifold or a template)
\cite{BW1, BW2}.
The spectrum of UPOs in the attractor
is a subset of all the orbits contained in the knot holder.
This subset is specified by a basis set of orbits \cite{Mindlin93a}.
The knot holder that describes an embedded chaotic attractor
is identified by extracting a rather small set of
orbits from the attractor and determining their
topological organization \cite{Gabo}.  As a result, knot holders
are invariant under diffeomorphisms isotopic to
the identity, since they are derived from the topological
indices of periodic orbits, which do not change under isotopy.
Knot holders can differ only by the indices that
describe the distinct equivalence classes of diffeomorphisms.
These are: the parity index $\pm 1$, the global torsion
$n$, and the knot type of the embedding into $R^3$,
including the framing index $f$.  Further, as the spectrum
of UPOs in a chaotic attractor is a diffeomorphism
invariant, every embedding of a chaotic attractor has
the same basis set of orbits.

A knot-holder
has as many branches as the number of symbols required to uniquely
identify the unstable periodic orbits in the
attractor.  This number, as well as the symbolic name of each periodic
orbit, can be determined by constructing a generating partition
\cite{GP1, GP2, GP5, GP6, GP7}.
Techniques have also been developed to construct the
knot-holder without prior knowledge of a symbolic encoding, by
searching directly for the simplest template with a set 
of orbits isotopic to the experimental one~\cite{ToC,GP3,GP4}. A
generating partition can  
then be constructed from this information \cite{GP3, GP4,GP4bis}.

A knot-holder has one or more branch lines.  Two or
more branches leave from each 
branch line (``stretching process''), and two or more branches meet 
at each branch line (``squeezing process'').  Since knot holders are
surrogates for chaotic attractors \cite{BW1,BW2}, we regard
information about which branches leave each branch line
and which meet at each branch line as describing the
mechanism generating chaos.

\begin{figure}   
\begin{tabular}{c}
\includegraphics[angle=0,height=7.0cm]{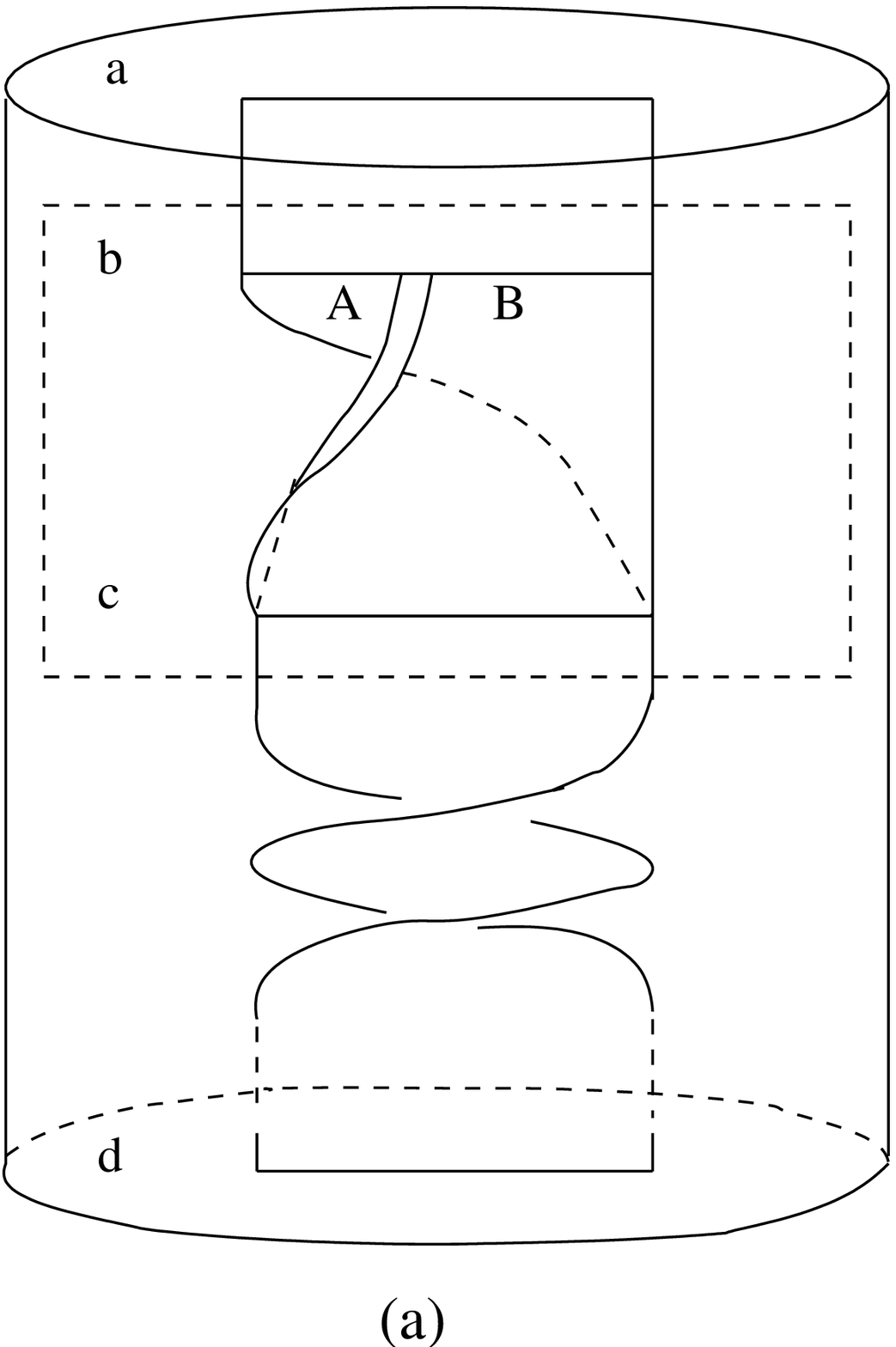}\\
 \\
\includegraphics[angle=0,width=7.0cm]{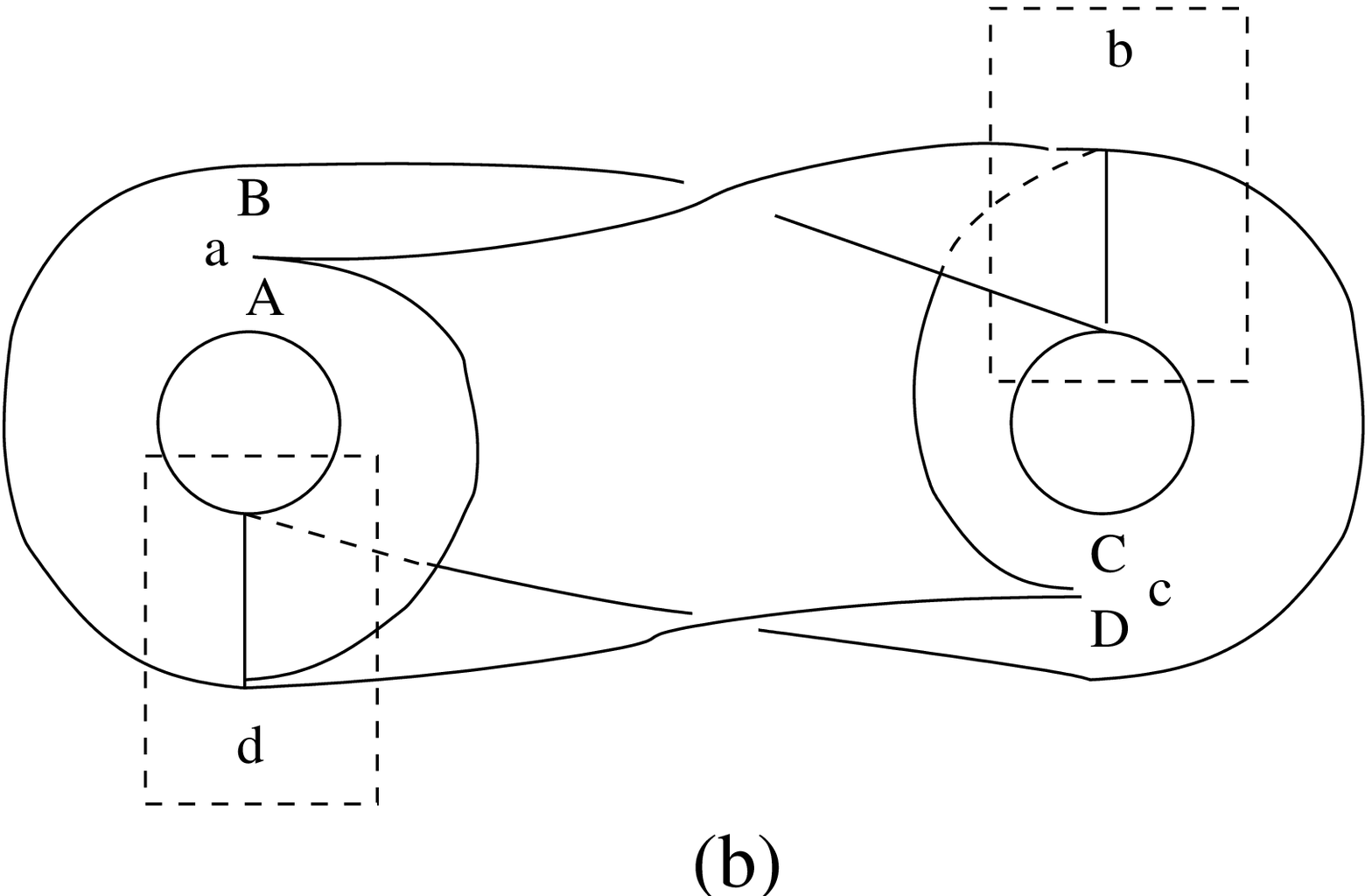}
\end{tabular}
  \caption{(a) Knot holder for the R\"ossler attractor,
shown inside a torus $D^2 \times S^1$.  The flow
enters at $a$, is split at $b$, joined at $c$, and
``leaves'' at $d$.  Periodic boundary conditions
identify $a$ and $d$.  We also identify $b$ with $a$ and 
$c$ with $d$.  
(b)  Knot holder for the Lorenz attractor,
shown inside a genus-three torus.  Branches $A$ and $B$
split at $a$ while $C$ and $D$ split at $c$.  Branches
$C$ and $B$ join at $b$ while $D$ and $A$ join at $d$.
In both cases the mechanism is shown within the dashed box
(a) or boxes (b).}
  \label{fig:knot-holders}
\end{figure}

Chaotic attractors in a torus (genus one) possess a single
branch line \cite{TTRG1}, which may be an 
interval (R\"ossler and Duffing
attractors) or a circle (van der Pol attractor).  For
attractors in $D^2 \times S^1$ by ``mechanism''
we mean explicitly the order
in which branches leave the branch line (left to right) 
or circle (clockwise or counterclockwise) 
and the order in which the
branches are squeezed together when they return to the
branch line (front to back) or circle (inside to outside)
\cite{classification}.  In the genus one case,
mechanism describes how the branch curve (line, circle)
is folded back into itself in one forward iteration.
The return flow, from the output side of the branch line
(lines $c$ to $d$ in Fig. \ref{fig:knot-holders}(a))
to the input side (lines $a$ to $b$ in Fig. 
\ref{fig:knot-holders}(a)) is assumed to
preserve order.  The ``mechanism'' is shown within the
dashed rectangle of Fig. \ref{fig:knot-holders}.  
The part of the branched
manifold describing the flow from $b$ to $c$ is the part
of the branched manifold that describes stretching
(the divergence of branches $A$ and $B$) and squeezing
(the joining of branches $A$ and $B$).  This is the part
of the branched manifold describing ``mechanism.''
This knot-holder has only one branch line.  We have
shown four in Fig. \ref{fig:knot-holders} to emphasize
the various roles played by that branch line.

\begin{figure}[htb]   
\begin{center}
\includegraphics[angle=0,width=7.0cm]{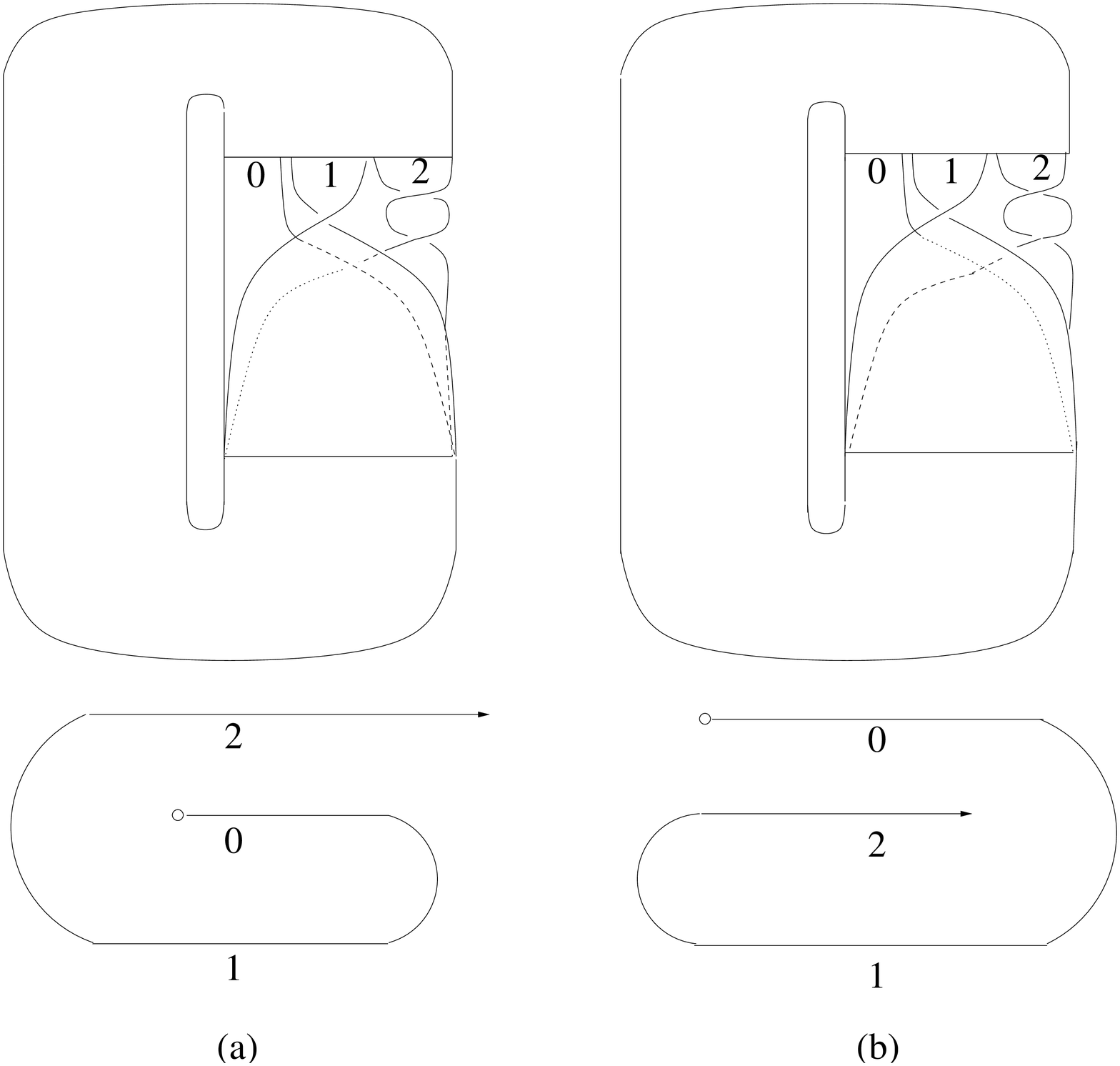}
\[ \begin{array}{ccccc}
 {\rm Branch} &  {\rm Matrices}& &{\rm Branch}  & {\rm Matrices}\\
 &  & & & \\
\begin{array}{c} 0 \\ 1 \\ 2 \end{array} &
\left[  \begin{array}{ccc}
0 & 0 & 2 \\
0 & 1 & 2 \\
2 & 2 & 2 \end{array} \right]& &
\begin{array}{c} 0 \\ 1 \\ 2 \end{array} &
\left[  \begin{array}{ccc}
0 & 0 & 0 \\
0 & 1 & 2 \\
0 & 2 & 2 \end{array} \right]\\
  & & & & \\ 
 & \left[  \begin{array}{ccc} 1 & 2 & 0 \end{array} \right] & & 
 & \left[  \begin{array}{ccc} 0 & 2 & 1 \end{array} \right]\end{array}
\]
\end{center}
\caption{Three-branch knot-holder for an (a) inside-to-outside
and (b) outside-to-inside ``jelly roll'' mechanism.  the template
matrices and arrays that describe these branched manifolds
algebraically are also shown.}
\label{fig:gateau_roule}
\end{figure}

Knot-holders for chaotic attractors in a genus-one torus 
are classified by a pair of matrices \cite{RMP,
ToC, RRR, Gabo}.  If $n$ symbols are required to label periodic
orbits, one matrix (the ``template matrix'') is an $n \times n$ matrix
and the other (``array matrix'' or ``joining matrix'') is a $1 \times
n$ matrix.  These two matrices are shown in Fig. \ref{fig:gateau_roule}
for two particular knot-holders.  One (Fig. \ref{fig:gateau_roule}(b))
is the outside to inside scroll template
with three branches, which has been observed in (embeddings of)
experimental data from
lasers \cite{lasers2,lasers,lasers3} and from neurons \cite{neurons}.
The other (Fig. \ref{fig:gateau_roule}(a)) is the inside-to-outside
gateau roul\'e.
The diagonal matrix elements $T_{ii}$ of the template matrix describe
the local torsion (measured in units of $\pi$) for branch $i$.  The
off-diagonal matrix elements $T_{ij}=2 \times Link(i,j)$ are twice the
linking numbers of the period-one orbits in branches $i$ and $j$.  The
array matrix describes the order in which the branches are glued
together at the branch line: the smaller the integer entry, the
further from the viewer in the projection.

Mechanisms that differ by being mirror images or by having integer
global torsion are represented by closely related matrices. In the
opposite parity case, the mirror image knot holder has all integer
entries with opposite signs. In the case of global torsion $n$, the
even integer $2n$ is added to all entries in the template matrix. The
matrices that describe these two variations of the gateau-roul\'e
mechanism [c.f. Fig. \ref{fig:gateau_roule}(b)] are

\begin{equation}  \begin{array}{ccc}
{\rm Branch} & {\rm Parity} &  {\rm Global~Torsion~}n\\
 & & \\
\begin{array}{c} 0 \\ 1 \\ 2 \end{array} &
\left[  \begin{array}{ccc}
0 & 0 & 0 \\
0 & -1 & -2 \\
0 & -2 & -2 \end{array} \right] & 
\left[  \begin{array}{ccc}
2n & 2n & 2n \\
2n & 2n+1 & 2n+2 \\
2n & 2n+2 & 2n+2 \end{array} \right]\\
  & & \\
 & \left[  \begin{array}{ccc} 0 & -2 & -1 \end{array} \right]& 
\left[  \begin{array}{ccc} 0 & 2 & 1 \end{array} \right] \end{array}
\label{global_torsion}
\end{equation}

Embeddings with nontrivial knot type do little to alter the matrices
that describe the mechanism that generates chaotic behavior
\cite{Aruna}. Nontrivial knot type may change parity and add global
torsion, depending on the framing \cite{Rolfsen} of the knot (see
Sec.~\ref{sec:embeddings}). 

\begin{figure}[htb]   
\begin{center}
\includegraphics[angle=0,width=3.5cm]{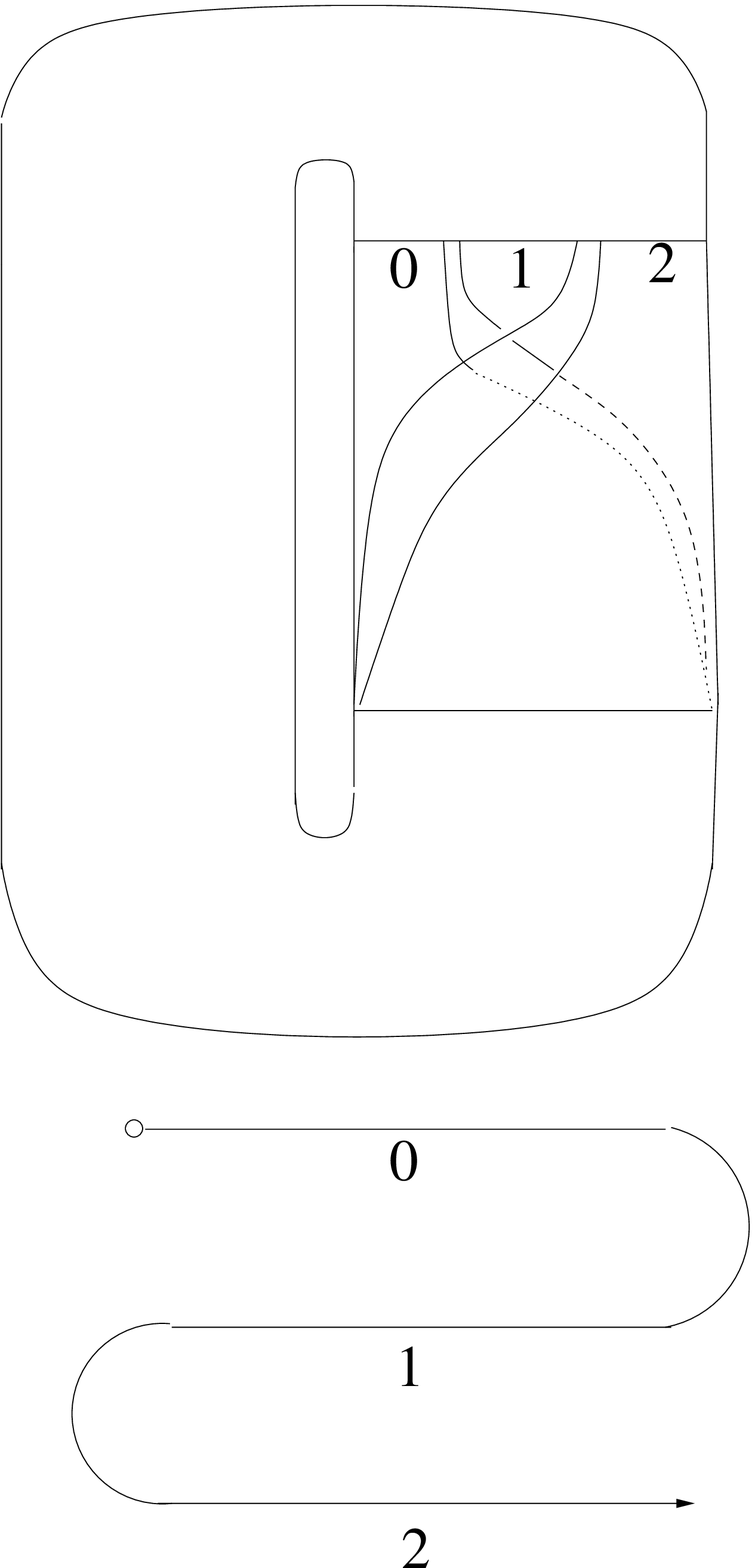}
\[  
\begin{array}{cc}
{\rm Branch} & {\rm Matrices} \\
 & \\
\begin{array}{c} 0 \\ 1 \\ 2 \end{array} & 
\left[  \begin{array}{ccc}
0 & 0 & 0 \\
0 & 1 & 0 \\
0 & 0 & 0 \end{array} \right]\\
  & \\
 &  \left[  \begin{array}{ccc} 0 & 1 & 2 \end{array} \right] \end{array}
\]
\end{center}
\caption{Three-branch knot-holder for an $S$-mechanism,
along with its template matrix and joining array.}
\label{fig:S_mechanism}
\end{figure}

To be explicit, a mechanism that generates chaos requiring three
symbols can be of two types: a scroll mechanism
(Fig.~\ref{fig:gateau_roule}) or an ``S'' mechanism. The template 
for the latter is shown
in Fig.~\ref{fig:S_mechanism}, along with its description
in terms of matrices.

If one embedding of data reveals a scroll
template, all embeddings will reveal a scroll mechanism.
If on the other hand one embedding reveals an $S$ mechanism,
every other embedding of these data will also reveal an
$S$ mechanism.  This is true because no transformation involving
sign changes or addition of global torsion 
[c.f., Eq. (\ref{global_torsion})] can change
the description given in Fig. \ref{fig:gateau_roule} to the description
given in Fig. \ref{fig:S_mechanism}.  The mechanism (scroll or $S$) is an
invariant of embeddings.  

Similarly, a horseshoe mechanism $\mathcal{H}(n,\epsilon)$ will be
described in all embeddings by template matrices
\begin{equation}  \begin{array}{cccc}
 & {\rm Branch} &~~~ & {\rm Matrices} \\
 & & & \\
\begin{array}{c}
{\rm Horseshoe~with}\\
{\rm Parity~and}\\
{\rm Global~Torsion} \end{array}&
\begin{array}{c} 0 \\ 1 \end{array} & &
\left[  \begin{array}{ccc}
2n & 2n \\
2n & 2n+\epsilon  \end{array} \right]\\
  & & & \\
 & & &
\left[  \begin{array}{ccc} 0 & \epsilon \end{array} \right] \end{array}
\label{horseshoe_mechanism}
\end{equation}
with $\epsilon=\pm1$ and $n$ indicating parity and global torsion
or framing index,
respectively. Matrices~\eqref{horseshoe_mechanism} describe all
possible templates with two branches folding over each other. 
A mechanism identified as a horseshoe in one embedding 
is a horseshoe in any embedding.

\section{Topological Indices}
\label{sec:topological}

Relative rotation rates are the natural topological index 
for periodic orbits in the torus $D^2 \times S^1$ \cite{RRR}.  
Linking numbers are the natural topological index 
for periodic orbits in $R^3$.

Assume $A$ and $B$ are two periodic orbits in some embedding
in the torus $D^2 \times S^1$, and that their relative rotation
rates are $R_{ij}(A,B)$.  These fractions are invariant under
diffeomorphisms isotopic to the identity.  Under diffeomorphisms
$D^2 \times S^1 \rightarrow D^2 \times S^1$ with global
torsion $n$, or with parity $-1$, that map $A \rightarrow A'$ and
$B \rightarrow B'$

\[   \begin{array}{lrclll}
{\rm Global~torsion}=& n & &R_{ij}(A',B') & = &R_{ij}(A,B)+n\\
{\rm Parity}=& -1 & & R_{ij}(A',B') & = &-R_{ij}(A,B)\end{array}
\]

Diffeomorphisms $D^2 \times S^1 \rightarrow R^3$ map
$A \rightarrow A''$ and $B \rightarrow B''$.  For these
closed curves in $R^3$ it is possible to compute both
relative rotation rates and linking numbers.  Under
the natural embedding \cite{RRR} (c.f., Fig. \ref{fig:embed_frame})
\[\begin{array}{lll}
R_{ij}(A'',B'') & \rightarrow &R_{ij}(A,B)\end{array}
\]
Under an embedding into $R^3$ with framing index $f$
\[\begin{array}{lll}
R_{ij}(A'',B'') & \rightarrow &R_{ij}(A,B)+f\end{array}
\]
In all cases the linking numbers  of $A''$ and $B''$ in 
$R^3$ are the sum of their relative rotation rates \cite{RRR}:
\[
L(A'',B'') = \sum_{i=1}^{p_A}\sum_{j=1}^{p_B}R_{ij}(A'',B'')
\]
where $p_A = p_{A''}$ is the period of orbits $A$ and $A''$, and
similarly for $B$.  The dependence of the linking number
of $A''$ and $B''$ on the framing index $f$ is
\[
L_f(A'',B'') = L_0(A'',B'')+ f p_A p_B 
\]
The framing index $f$ for embeddings
$D^2 \times S^1 \rightarrow D^2 \times S^1 \subset R^3$
can be considered, for all practical purposes,
as equivalent to the global torsion $n$ for embeddings
$D^2 \times S^1 \rightarrow D^2 \times S^1$.

\section{Perestroikas}
\label{sec:perestroikas}

Up to this point the discussion has concentrated on
embeddings of a single attractor.  Usually experiments
that generate chaotic attractors are carried out
over a range of control parameter values in an effort
to create the equivalent of a bifurcation diagram.
In this section we discuss fixed embeddings of a
family of attractors and the dual process: families of
embeddings of a single attractor.

The first topological analysis of a family of
chaotic attractors was carried out in \cite{Papoff}.
A single embedding was used to analyze many data sets
from lasers with saturable absorbers operated
with three different absorbers and under various operating
conditions.  This analysis revealed that through all
these changes the underlying branched manifold never
changed: it was only the basis set of orbits that changed
\cite{Papoff,Mindlin93a}.
Results for an $NMR$ laser \cite{Tufillaro91a} and a nonlinear
vibrating string \cite{Tufillaro_string} were the same.
Subsequently, studies of the periodically driven
Duffing oscillator \cite{Gilmore95a}, $CO_2$ lasers with modulated
losses \cite{GP3,Lefranc93a}, an $Nd$-doped YAG laser\cite{lasers2}, 
an $Nd$-doped fiber laser \cite{lasers,lasers3}, 
and sensory neurons \cite{neurons} showed that
the underlying branched manifold was a ``gateau roul\'e''
or ``jelly roll'' branched manifold \cite{RMP, ToC}, 
and that under variation of the  modulation frequency 
the flow was directed to branches of this branched 
manifold with systematically increasing torsion.

In light of the results presented
in the preceeding sections, these conclusions are 
embedding-independent: they would have been reached 
using any embedding.
First, the variation of torsion with control parameters was observed
using a fixed 
embedding, hence is due to physical effects and not to the choice of
embedding. Within a fixed-embedding study, the standard horseshoe
$\mathcal{H}(0,1)$ is topologically distinct from a ``reverse''
horseshoe $\mathcal{H}(1,-1)$ (as observed in~\cite{lasers2}).
Second, the spiral structure that globally describes
attractors observed at different control parameters values would not
have been affected if an embedding with different knot type, torsion
and parity had been chosen.

It is often the case that families of mappings are
studied in an effort to identify an optimum embedding.
The method of minimum mutual information \cite{Fra86} 
was developed for
precisely this reason.  The first systematic study of 
the way the topological properties of an embedded
attractor can depend on the embedding, or change with
the embedding parameters, was carried out in \cite{Aruna}.
This is the example (2) summarized in the Introduction.
Mappings with a delay $\tau < \tau_1$ provided embeddings,
as did mappings with $\tau > \tau_2$.  In both cases,
changing the delay $\tau$ by a little had no effect
on the topological indices of the periodic orbits.
In both cases the torus embedded in $R^3$ wound around
a vertical axis three times before closing.
In the transition from one regime of
embeddings to the other all relative rotation
rates changed by $ \pm 2$ (depending on whether the
time delay $\tau$ increases or decreases).  
In the interval $\tau_1 < \tau < \tau_2$
the mapping exhibited self intersections, of the type
indicated by the arrows in Fig. \ref{fig:embed_frame}(c).
The knot type of the embedding into $R^3$ remained unchanged
but its framing in $R^3$ changed.  Further, the change was
by an even integer.  This is a signature for
framing changes caused by change in handedness
of writhe \cite{ToC}.

\section{Summary and Conclusions}
\label{sec:summary}

When a low dimensional chaotic attractor is embedded in a
three dimensional space, its topological properties
depend on the embedding.  We show that, for a large class
of low dimensional attractors there are three 
topological properties that are embedding-dependent and 
one that is embedding-independent.  The embedding-dependent
properties are: parity, global torsion, and knot type.  In the
latter case (of mappings $D^2 \times S^1 \rightarrow R^3$),
the framing index is the global torsion.  The
embedding-independent property is the mechanism that
acts in phase space to create the chaotic attractor.
Mechanism is defined in Fig. \ref{fig:knot-holders}
in terms of branched manifolds.
The class of chaotic attractors for which these results
hold includes all genus-one attractors: those whose phase space is 
equivalent (diffeomorphic)
to a torus $D^2 \times S^1$.  This class includes
the R\"ossler attractor, periodically driven two-dimensional
nonlinear oscillators such as the Duffing, van der Pol, and
Takens-Bogdanov attractors, and most of the experimentally
generated chaotic attractors that have been studied
by topological methods.   The principal result is that 
any single embedding of a three dimensional attractor 
in this class suffices to determine the
mechanism that has generated the chaotic data.
This class does not include the Lorenz
attractor and other attractors with more than one
``hole in the middle.'' 

\appendix
\section{Classification of embeddings of $D^2 \times S^1$ into
  $R^3$.}
\label{sec:app}
\subsection{Introduction}

In this appendix, we provide the interested reader with more details
about how embeddings of genus-one attractors can be classified in
terms of knot type, torsion and parity.

Assume that two embeddings $\Psi_1$ and $\Psi_2$ of a chaotic
attractor are possible.  The simplest case is when $\Psi_1$ and
$\Psi_2$ are isotopic: one embedding can be deformed continuously into
the other. Equivalence of the topological properties of the two
embeddings then trivially follows from the invariance of the
topological indices of closed curves with respect to smooth
deformations that do not induce self-intersections.

When $\Psi_1$ is not isotopic to $\Psi_2$, we exploit the assumption
that the original strange attractor can be enclosed in a genus-one
torus. We first note that a diffeomorphism (or homeomorphism) mapping
the original attractor to a reconstructed attractor is defined on
neighborhoods of these two strange sets, and can easily be extended to
a diffeomorphism (or homeomorphism) between solid tori contained in
these neighborhoods and enclosing the attractors.

Since isotopic embeddings are equivalent, determining how topological
properties of two genus-one embeddings of an attractor can differ thus
simply amounts to studying isotopy classes of embeddings of $D^2\times
S^1$ into $\mathbb{R}^3$. There are two levels in the classification
of these isotopy classes, because there are two ways in which two
embedded solid tori can be non-isotopic. The first level is extrinsic
and deals with how the core of the solid torus is embedded in
$\mathbb{R}^3$. When shrunk to their respective cores, two solid tori
are isotopic if they have the same knot type. The second level is
intrinsic and deals with how torus boundaries $\partial (D^2 \times
S^1) = T^2$ are mapped to torus boundaries. Two embeddings such that
the cores of the embedded tori have the same knot type can still be
non-isotopic if the homeomorphism mapping the boundary of one torus to
the boundary of the other is not isotopic to identity.  

From this classification, we finally conclude that there are three
degrees of freedom in which two embeddings of a genus-one attractor
into $R^3$ can differ: knot type (extrinsic level), torsion and parity 
(at both extrinsic and intrinsic levels).

\subsection{Extrinsic level: knot type}

A necessary condition for two embeddings of a manifold $M$ to be
isotopic is that their restrictions to a given submanifold $M'\subset
M$ are isotopic \cite{Rolfsen}. In particular, consider the core of
the solid torus $D^2\times S^1$, i.e., the submanifold
${\mathcal{C}}=\{B\}\times S^1$ with a base point $B\in D^2$. Two
isotopic embeddings of $D^2\times S^1$ into $\mathbb{R}^3$ must embed
$\mathcal{C}$ into $\mathbb{R}^3$ isotopically. Since $\mathcal{C}$ is
an embedding of $S^1$ into $D^2\times S^1$, embeddings of $C$ in
$\mathbb{R}^3$ can be classified as embeddings of $S^1$ in
$\mathbb{R}^3$, i.e., as ordinary knots. Two torus cores are thus
isotopic if and only if they have the same knot type. Conversely, two
embedded solid tori whose cores are knotted in different ways cannot
be isotopic.

\begin{figure}[htbp]
  \centering
\includegraphics[angle=0,width=3.2in]{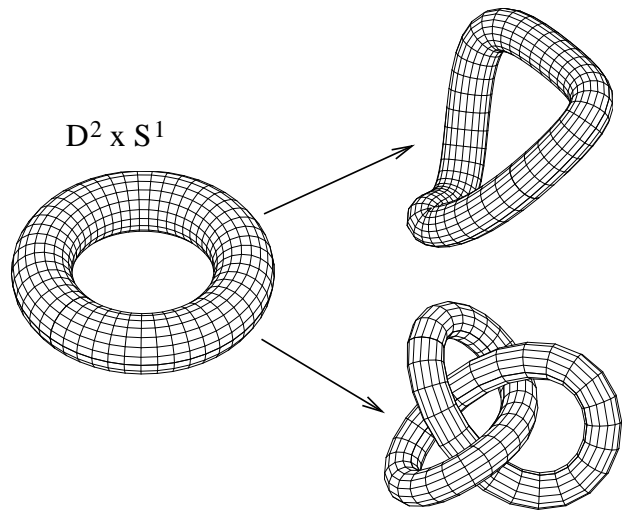}
  \caption{Two embeddings of $D^2\times S^1$ as solid tori in
    $\mathbb{R}^3$ cannot be isotopic if their cores are not isotopic.}
  \label{fig:isotopycores}
\end{figure}

The actual situation (embedding) of tori in $\mathbb{R}^3$, as
described by the knot type of the torus core, is called the extrinsic
structure \cite{Weeks}. Every (tame) knot can be used as a centerline
for a torus that is embedded in $\mathbb{R}^3$.

Assume that two embedded solid tori have isotopic cores. This
allows us to superimpose the boundaries of the two solid tori by
isotopy deformations. This does not imply that the two embeddings are
isotopic. However, this indicates that we can now study the
classification of embeddings at an intrinsic level, by considering
mappings of the solid torus into itself and forgetting about position
in $\mathbb{R}^3$ (looking now at the torus from the inside rather
than from the outside). Thus, knot type of the torus core captures all
information about isotopy classes at the extrinsic level.

\begin{figure}[htbp]
  \centering
\includegraphics[angle=0,width=3.2in]{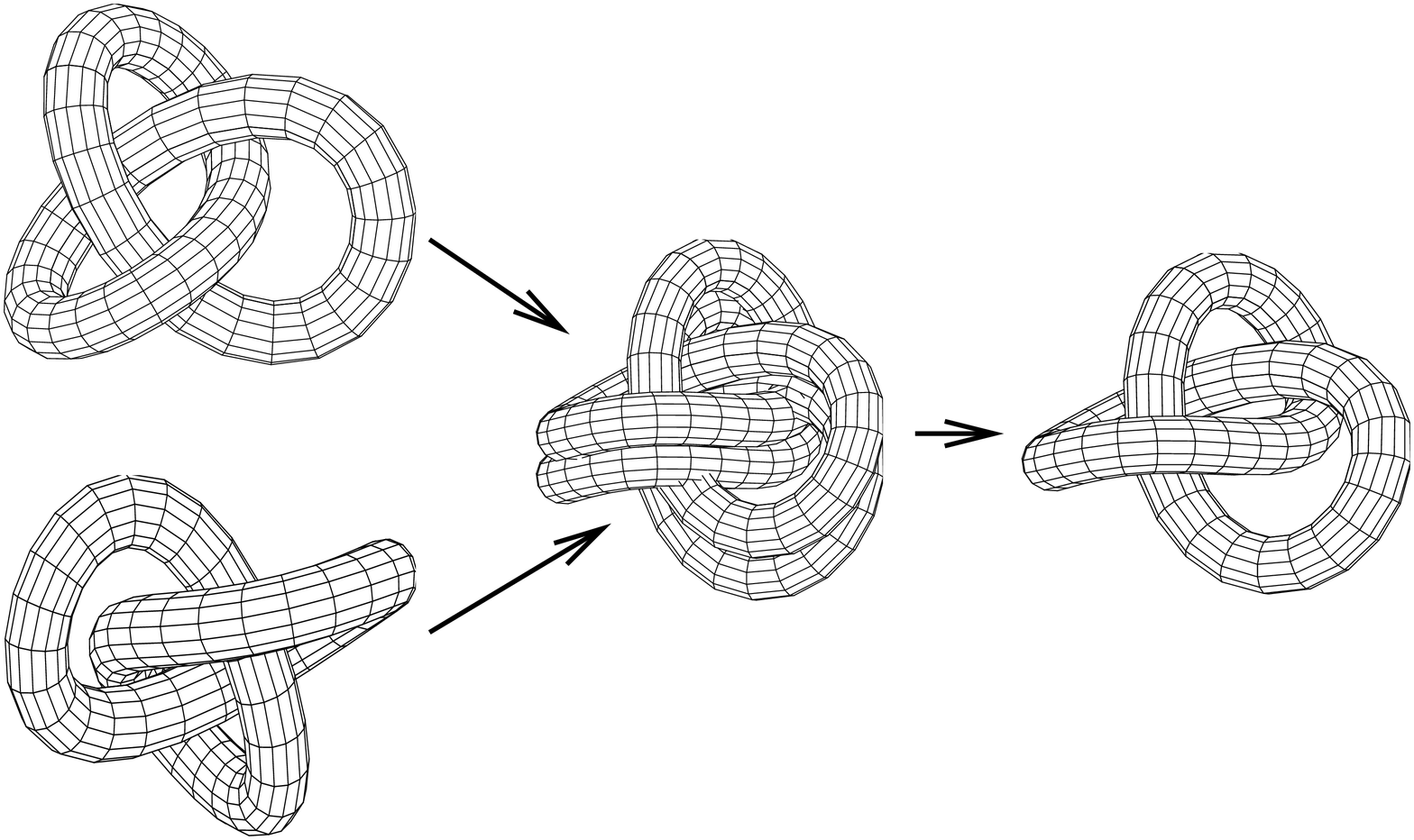}
  \caption{The boundaries of two  embedded solid tori with isotopic
    cores can be superimposed by isotopy deformations.}
  \label{fig:isotopyboundaries}
\end{figure}

\subsection{Intrinsic level: global torsion and parity}

We consider now two embeddings $\Psi_1$ and $\Psi_2$ of $D^2\times
S^1$ such that the cores of the embedded solid tori are isotopic and
their boundaries are superimposed (Fig.~\ref{fig:isotopyboundaries}).
A necessary condition for the two embeddings to be isotopic is that
their restriction to the boundary of the tori are isotopic, as with
any submanifold. This condition is equivalent to requiring that the
restriction of $(\Psi_1)^{-1}\Psi_2$ to the boundary of the first
torus is isotopic to identity. It is also a sufficient condition
because a homeomorphism of the torus that has its restriction
isotopic to identity is isotopic to identity~\cite{Rolfsen}. Thus, we
are left with studying isotopy classes of homeomorphisms of the
boundary $\partial (D^2 \times S^1) = T^2$ into itself.

The group of homeomorphisms of this surface to itself modulo
isotopically equivalent embeddings is called the mapping class group
and is equivalent to the modular group of $2 \times 2$ matrices
$GL(2;Z) = \left[ \begin{array}{cc} a & b \\ c & d
  \end{array} \right]$, with $a,b,c,d$ integer and $ad-bc = \pm 1$
\cite{Rolfsen}.  This group describes how closed curves $S^1 \subset
T^2$ are mapped to closed curves in $T^2$ under the homeomorphism. The
description is given in terms of the basis set of (two) loops for the
homotopy group of $T^2$.  These two cycles are the meridian and the
longitude.  The meridian can be regarded as the small loop that goes
around a tire ``the short way'' and the longitude as a long loop that
goes around the tire ``the other way'' (c.f., Fig.
\ref{fig:tori}(a)).  At the topological level (homeomorphism) they
are more or less equivalent. At the level of dynamical systems they
are not.  The meridian bounds a disk that lies inside $D^2 \times S^1$
and can be taken as everywhere transverse to the flow that generates
the strange attractor or its embedding.  This disk can be chosen as a
global Poincar\'e surface of section.  The longitude can be chosen in
the direction of the flow.  By restricting to the topology underlying
the dynamics, we investigate the class of inequivalent homeomorphisms
of the torus boundary into itself.  These are 
described by modular group operations of the form
\begin{equation}
M=\left[
  \begin{array}{cc} 1 & n \\ 0 & \epsilon \end{array}
\right]\label{eq:dynmodular}
\end{equation}
with $\epsilon=\pm 1$. We interpret the integer $n$ as the number of
times the longitude links the core of the solid torus $D^2 \times
S^1$.  Dynamically, $n$ is the global torsion of the embedding.  The
sign $\epsilon=\pm1$ identifies the parity of the torus homeomorphism.

A point has to be made regarding parity. The mirror image of an
embedding is also an embedding, which differs from the original
embedding only by orientation. In the mirror image of an embedding,
all the topological invariants are multiplied by $-1$. Since an
embedding and its mirror image cannot be isotopic because orientation
is preserved under isotopy (orientation cannot change without inducing
self-intersections at some stage of the deformation), parity has to be
taken account when classifying embeddings of solid tori into
$\mathbb{R}^3$.

Mirror image transformations act both at the extrinsic and intrinsic
levels. Their action at the extrinsic level is easily incoporated in
the the knot type, which is changed into its mirror image. At the
intrinsic level, parity is taken into account through the sign of the
lower diagonal entry $\epsilon$ of the modular transformation
\eqref{eq:dynmodular}, which is also its determinant.

\subsection{Summary}

Embeddings of genus-one attractors can be classified by studying
isotopy classes of the cores of the embedded tori and of their
boundaries. There are three degrees of freedom by which embeddings of
genus-one attractors into $\mathbb{R}^3$ can differ: knot type of
extrinsic embedding, global torsion, parity (handedness).

\end{document}